# Nanoscale textured superconductivity in Ru-substituted BaFe$_2$As$_2$ : a challenge to a universal phase diagram for pnictides

Y. Laplace,[1] J. Bobroff,[1] V. Brouet,[1] G.Collin,[1] F. Rullier-Albenque,[2] D. Colson,[2] A. Forget [2]

[1]*Laboratoire de Physique des Solides, Univ. Paris-Sud, UMR8502, CNRS, F-91405 Orsay Cedex, France*
[2]*Service de Physique de l'Etat Condensé, Orme des Merisiers, CEA Saclay (CNRS URA 2464), 91191 Gif sur Yvette cedex, France*

$^{75}$As NMR experiments were performed in Ba(Fe$_{1-x}$Ru$_x$)$_2$As$_2$ for x=0 to 80%. Magnetic fractions and NMR lineshapes demonstrate that Ru substitution destroys the antiferromagnetic (AF) order inhomogeneously with a magnetic moment distributed from 0.9 to 0 $\mu_B$. Superconductivity emerges at intermediate Ru doping and coexists with AF order only in the regions where moments are smaller than ~0.3 $\mu_B$, resulting in an original nanoscale texture. This situation contrasts with that of Co substitution, challenging the apparent universality of the phase diagram in Fe-based superconductors.

Cuprates and Fe-based superconductors [1], the two known high T$_C$ superconductors, display a similar phase diagram where the destabilization of an antiferromagnetic (AF) order leads to superconductivity. This is obtained only by charge doping in cuprates, while it can be achieved through very different means in Fe-pnictides: not only charge doping by heterovalent substitution, but also isovalent substitution or hydrostatic pressure [2]. How such different parameters produce apparently similar physics is an open question at the heart of the physics of Fe-pnictides.

In the archetype Fe-pnictide parent compound BaFe$_2$As$_2$ - a compensated semi-metal [3,4] - the AF order is attributed to a spin density wave resulting from a good nesting between the hole and electron Fermi surfaces at the AF wave vector [5,6], even though this picture is still debated [7]. Charge doping produced for example by Co substitution at Fe site induces an increase of the size of the electron pocket [4,8], hence a destabilization of the nesting condition and a destruction of the antiferromagnetism [6], allowing for superconductivity to develop. On the contrary, with an isovalent substitution like Ru at Fe site, the nesting condition remains good due to the compensated electronic structure [9]. But surprisingly, Ru is found to destroy the AF order and lead to SC as well, with even the same ordering temperatures T$_N$ and T$_C$ as Co (upper panel of fig.1) [10-12]. The reason why isovalent Ru and heterovalent Co produce similar effects is still unclear. It is essential to capture at a local scale the physics at play nearby each Ru atom to understand this paradox. This also raises the issue of the effect of in-plane substitution in these materials.

Previous studies in Ru-substituted BaFe2As2 have mainly been performed through macroscopic probes [11-13]. Nuclear Magnetic Resonance (NMR) should help to settle this issue, as it measures the AF or SC states at local scale, but the only reported study in Ba(Fe$_{1-x}$Ru$_x$)$_2$As$_2$ did not focus on the ground state local properties [14]. We present an NMR study in the low temperature regime of Ba(Fe$_{1-x}$Ru$_x$)$_2$As$_2$ which allows to understand how Ru destroys antiferromagnetism and leads to superconductivity. The mechanism is very different from electronic Co doping and occurs at the nanometer scale through local effects of Ru in its immediate vicinity. The resulting ground state mixes superconductivity and magnetism in a pretty unique way. This result puts into question the "universality" of the phase diagram.

Polycrystalline samples of Ba(Fe$_{1-x}$Ru$_x$)$_2$As$_2$ were synthesized by solid-state reaction using small pieces of Ba metal and powders of Fe, Ru and As. Stoichiometric mixtures loaded in alumina crucibles were sealed in evacuated quartz tube and calcined at 975°C (925°C for x=0) during 36h. Rietveld analysis revealed that all samples were single phase excepted for x=0.80 where ~2% of Ru and RuAs$_2$ were detected after the second annealing. Upon Ru substitution, *a* crystallographic parameter increases linearly while *c* decreases in agreement with Ref.[11]. SC fractions were measured by diamagnetic shielding using a SQUID for an applied field H=10G after Zero Field Cooling. $^{75}$As NMR was measured with standard Fourier Transform recombination techniques in fields ranging from H=7 T to 13 T. Intensity measurements versus temperature were systematically performed to check that the full signal is observed at all compositions.

NMR and SQUID measurements detailed hereafter allowed us to determine the magnetic and superconducting fractions reported in fig.1.

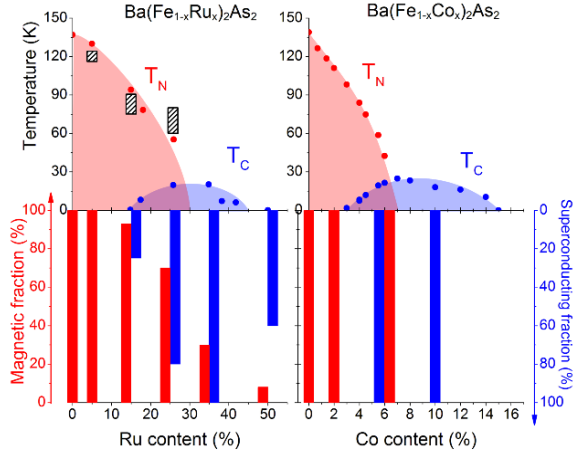

Fig. 1: Top : Phase diagrams established from transport measurements (red and blue dots) for Ru (left) and Co (right) substitutions [10,11]. Dashed boxes indicate the expected percolation transition (see text) Bottom : Evolution of the magnetic and superconducting volume fractions at low temperature (the latter being corrected from demagnetization factor of 1/3 to account for the Ru more or less spherical grain shapes) (for Co, fractions extracted from Ref.[15-17]).

In the Co doped compounds, the fractions are either 100% or 0%, indicating that whenever magnetism or superconductivity occurs, this happens over the whole sample homogeneously. These phases can even coexist together homogeneously on an atomic scale around $x_{Co}$=6% [17,21]. On the contrary, in the Ru substituted compounds, superconductivity develops at the expense of magnetism leading to intermediate fractions. One could suppose that AF and SC states just segregate spatially, i.e. never occur on the same Fe/Ru sites and that Ru compounds are trivially not homogeneous. Then one would expect the sum of the two fractions not to exceed 100%. This is clearly not the case for example at $x_{Ru}$=25%, where 70% of the sample volume is AF while 80% is SC. Furthermore, X-ray data show that the samples are single-phase. We are facing a subtle situation where AF and SC do not completely exclude each other spatially, but do not fully coexist as well.

Typical low temperature $^{75}$As NMR powder spectra are displayed in fig.2. At large Ru contents (Ru50%), the spectrum consists in the superposition of a narrow line and a broad triangular-shape background. Because $^{75}$As nuclear spin I=3/2 is sensitive to the electric field gradient (EFG), this results in a splitting of the NMR spectrum into the narrow line (transition $-\frac{1}{2} \leftrightarrow \frac{1}{2}$) and the background (quadrupolar satellite transitions $\pm\frac{3}{2} \leftrightarrow \pm\frac{1}{2}$). The latter is due to the fact that the EFG varies with orientation in the powder. Its triangular shape originates from the local modification of the EFG value due to Ru substitution.

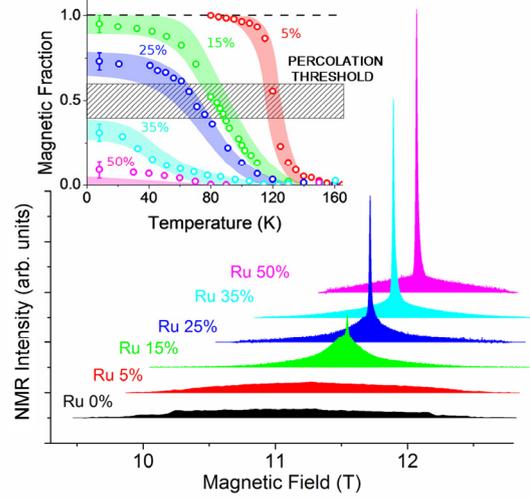

Fig. 2: NMR powder spectra for different Ru contents at low temperature (T<$T_N$ for 0% and 5% and T=8K otherwise). Inset: Temperature dependence of the paramagnetic fraction measured by integration of the spectral weight of the central line: experiment (circles), simulations (envelopes) and the percolation threshold for the magnetic ordering.

We were indeed able to simulate it using a large distribution of EFGs corresponding to $\nu_Q$ quadrupolar parameter ranging over a few MHz. The central line is very narrow because it is sensitive to the EFG only to second order in perturbation. Its position allows us to determine the NMR shift K proportional to the intrinsic magnetic susceptibility of (Fe/Ru) layers. We measured a shift K versus temperature identical for all Ru contents as in Ref.[14]. This demonstrates that Ru substitution does not change the doping in Fe layers contrary to electron or hole dopings which modify K linearly as a function of the substitution content [18,19]. Ru is indeed isovalent to Fe and does not unbalance the electron-hole compensated semi-metal, as also confirmed by ARPES measurements [9].

In the undoped parent compound BaFe$_2$As$_2$ (Ru0%), the central line is absent and the broad component has a more rectangular shape with sharp edges. This signals an AF magnetic order which induces large internal fields that distributes the position of both central and satellite transitions over a few Tesla. The shape can be well simulated assuming the stripe AF order found in neutron experiments [5] using hyperfine couplings of $^{75}$As with Fe moments [16]. When increasing Ru content and spanning the entire phase diagram from the full AF to the full paramagnet, we observe a progressive increase of the narrow line intensity relative to the background.

This indicates the growth of non frozen paramagnetic domains coexisting with the AF state.

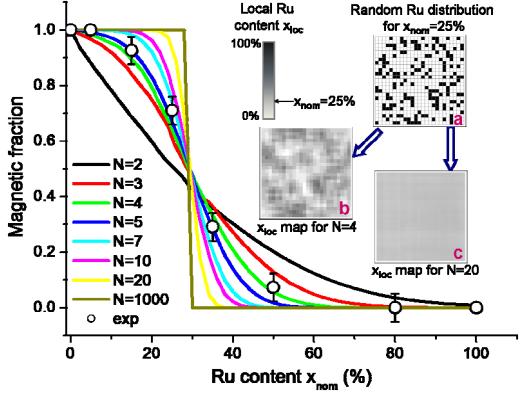

Fig.3: Magnetic fraction versus Ru nominal content: experiment (circle, Ru 0% and 100% from Refs.[14,16]) and simulations (solid lines) using a local Ru content model. Inset: (a) random distribution of Ru (black dots) in a layer of Fe sites (only a portion of size 25*25 of this layer is shown); a moving N*N averaging of this layer leads to (b) for N=4 and (c) for N=20.

In the inset of fig.2, we plot the AF volume fraction deduced from the wipeout of the central line spectral weight corrected from temperature effects [20]. This determination is independent of the precise analysis of the background itself. At very small Ru content, such as Ru5%, it monitors a sharp transition to a full AF order as expected. For intermediate compositions (Ru15%-35%), the ordering transition is much broader and paramagnetic fractions remain even at low temperature, establishing the coexistence of paramagnetic and AF domains in the ground state of the material, as reported in fig.1. But as already stressed, AF and SC do not just segregate spatially but partially coexist.

To understand the origin of this non-trivial coexistence, we propose a rough model. We assume that even if samples are single-phase with nominal content $x_{nom}$, Ru substitution does not have a homogeneous effect on electronic properties but a more local one. The Fe layers properties are assumed to be governed by the local Ru content $x_{loc}$, which differs from $x_{nom}$ if averaged over a small scale. Regions with $x_{loc}$ are then supposed to become AF at the $T_N(x_{loc})$ displayed in fig 1, even for regions with nanometer size. As an example, in fig.3a we plot the Ru randomly substituted at Fe sites for $x_{nom}$=25%. We compute the local Ru content $x_{loc}$ using a moving average over sub-units of N*N unit-cells. Small N leads to a large distribution of the Ru local content (fig.3b) and the coexistence in the same sample of AF islands for small $x_{loc}$ and paramagnetic ones for large $x_{loc}$. At large N, one recovers a homogeneous ground state either 100% or 0% AF with $x_{loc} \approx x_{nom}$ (fig.3c). This statistical model allows us to compute - with no adjustable parameter - the expected magnetic fraction as a function of Ru nominal content for various N (main panel of fig.3).

Our experimental results are well fitted with N=4-5 for all the Ru concentrations. This corresponds to AF domains with an area in between 1nm² and 2nm². With no additional parameter, we are able to compute the expected temperature dependence of the magnetic fraction, which is plotted in the inset of fig.2 on top of the experimental data. The envelope encodes all the uncertainties coming from the error bar over N and over $T_N(x_{loc})$. The agreement is fairly good and explains why the magnetic transitions are so broad: they originate from the large range of $T_N$ due to the distribution in local Ru content.

We can further check this model for $x_{nom}$=15% using now the information contained in the NMR spectral shape. The spatial distribution of $x_{loc}$ should result in a spatial distribution of the local moment amplitudes on Fe/Ru sites, hence on the internal fields probed by the $^{75}$As NMR. Following neutron experiments [13], we assume the same AF order as in the parent compound but with the moment m($x_{loc}$) varying linearly with $x_{loc}$ from m($x_{loc}$=0%)=0.9 $\mu_B$ to m($x_{loc}$=30%)=0.0 $\mu_B$. Using the $^{75}$As hyperfine couplings of Ref.[16], the expected NMR spectrum is computed for different N using the same spatial simulated distributions of $x_{loc}$ as exemplified in fig.3. The simulated powder spectra are shown in the upper panel of fig.4 for $x_{nom}$=15%. The experimental spectrum is well fitted for N=4, the very same value found previously from the independent analysis of the narrow line intensities. The magnetic texture for N=4 is displayed on the lower panel of fig.4. For larger Ru contents, similar spectrum simulations are less conclusive because quadrupolar transitions of the paramagnetic domains convolute with the AF component. But the main conclusions of our study based on the analysis of the magnetic and SC fractions are valid for all Ru contents independently of the spectrum shape.

We now turn to the establishment of superconductivity and the nature of its coexistence with magnetism. As pointed out earlier, fig.1 indicates that superconductivity must coexist with magnetism on some atomic sites but not on all of them. This can be understood in light of the Co case, where homogeneous coexistence between SC and AF was found in a range of dopings around x~5 to 6% when the AF moments are small enough compared to the parent compound [17, 21]. At these doping levels, the ordered moments were indeed found smaller than about 0.3$\mu_B$ [22]. We may expect SC to coexist with AF for Ru only in the small moments regions too. Assuming the same typical threshold of about $0.3\mu_B$ moments below which SC can develop, we can then simulate the expected texture, as displayed for the 15% Ru case on fig.4b. It consists in AF-only regions with large moments (red) and coexisting AF and SC regions with small moments (encompassed in blue). In this example, about 25% of the sample should be SC, which is compatible with the observed SC fraction in fig.1. A more accurate comparison with the experimental SC fraction is beyond the scope of this study, since it involves many other factors hard to estimate such as proximity effects or Josephson couplings between the SC islands.

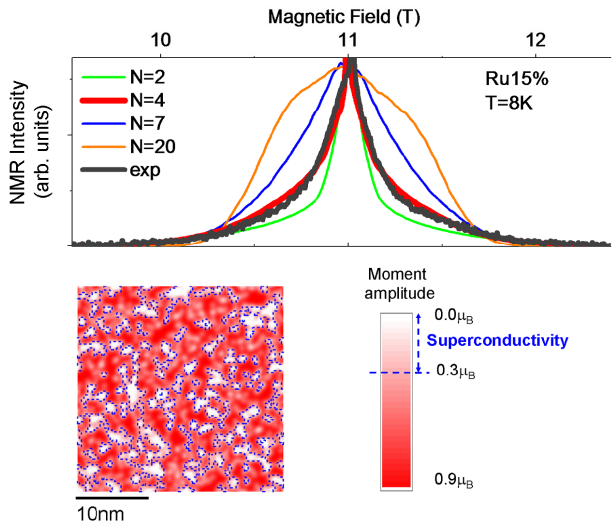

Fig. 4: <u>Top:</u> NMR experimental spectrum for Ru15% (dark gray) versus simulations (colored) done with the magnetic textures computed for N=2,4,7 and 20. <u>Bottom:</u> Typical distribution of the AF magnetic moment amplitude for N=4 in a 100*100 Fe layer. Blue lines encompass areas where SC and AF coexist.

How can we reconcile this inhomogeneous state for Ru with the apparent homogenous universal phase diagram of fig.1 ? The answer lies in the difference between the way resistivity and NMR probe this state. NMR is sensitive to any variation at the nanometer scale, while transport measurements used to determine the phase diagrams are not. In the Ru textured state, an AF (or SC) transition will be observed with resistivity measurement when the AF (or SC) islands percolate. This happens at two dimensions when about at least 50% of the sample volume is AF (or SC). Using the AF magnetic fraction of the inset of fig.2, we plot on fig.1 where this percolation should occur (dashed rectangles). It is indeed found very close to the resistivity determination of $T_N$. This could also explain why resistive transitions appear somewhat broader than in the homogeneous case of Co, as observed in Ref.[12]. Other macroscopic probes should also average out the Ru nanoscale inhomogeneities. Neutron scattering experiments in similar compounds do reveal an apparent ordered homogeneous AF state [13]. X-Ray crystallographic methods also detect a single homogeneous phase in our samples because they average over any nanoscale Ru induced distortion. This averaging was demonstrated in the chemically pressurized $FeSe_{0.5}Te_{0.5}$, where standard X-Rays detect a single environment for Fe despite the fact that FeSe and FeTe interatomic distances are different [23].

Why does Ru destroy AF order and allow SC to develop ? It has been argued that Co doping destabilizes the AF order in the reciprocal space by weakening the nesting between electron and hole pockets [6]. Here, Ru destroys the AF order in the real space instead, with a local effect. The physical mechanism for this weakening could be the local decrease of correlations between electrons around the Ru site due to more extended Ru orbitals as compared to Fe. This real space picture is qualitatively consistent with a magnetic dilution scenario as proposed for LaFeRuAsO [24]. Another scenario could be a local distortion modifying the chemical bonds with the surrounding ligand atoms (As in that case) that should have a drastic impact on electronic properties [25]. In both scenarios, a large content of Ru is needed to induce sizeable changes as observed here. Local structural studies of the Ru local environment would be very helpful, together with theoretical studies aimed at describing inhomogeneous electronic phases.

Note that Tc's are surprisingly similar between Ru and Co compounds, even though one would expect the Ru nanoscale SC to be weaker because of the need to establishing global phase coherence between the spatially disjoint SC islands [26]. We conclude that the SC state of the iron based superconductors demonstrates an unexpected high robustness against electronic inhomogeneity which remains to be fully understood.

In conclusion, the universality of the phase diagram of iron-based superconductors lies in the development of SC when the AF order is sufficiently destroyed. But the way this destruction is achieved is not universal. It can be done either with a mechanism acting in the reciprocal space, resulting in a homogeneous electronic state in the real space, or with a mechanism acting in the real space, leading to an intrinsically inhomogeneous electronic state on a local scale.

We acknowledge for fruitful discussions H. Alloul, F. Bert, S. Biermann, R. De Renzi, M. Marsi, P. Mendels and C. Simon, the ANR Pnictides and the RTRA Triangle de la Physique for support.